# Prostate Lesion Detection and Salient Feature Assessment Using Zone-Based Classifiers

2020 Summer STEM Institute

*Haoli Yin*

## Abstract

*Multi-parametric magnetic resonance imaging (mpMRI) has a growing role in detecting prostate cancer lesions. Thus, it's pertinent that medical professionals who interpret these scans reduce the risk of human error by using computer-aided detection systems. The variety of algorithms used in system implementation, however, has yielded mixed results. Here we investigate the best machine learning classifier for each prostate zone. We also discover salient features to clarify the models' classification rationale. Of the data provided, we gathered and augmented T2 weighted images and apparent diffusion coefficient map images to extract first through third order statistical features as input to machine learning classifiers. For our deep learning classifier, we used a convolutional neural net (CNN) architecture for automatic feature extraction and classification. The CNN results' interpretability was improved by saliency mapping to understand the classification mechanisms within. Ultimately, we concluded that effective detection of peripheral and anterior fibromuscular stroma (AS) lesions depended more on statistical distribution features, whereas those in the transition zone (TZ) depended more on textural features. Ensemble algorithms worked best for PZ and TZ zones, while CNNs were best in the AS zone. These classifiers can be used to validate a radiologist's predictions and reduce inter-reader variability in patients suspected to have prostate cancer. The salient features reported in this study can also be investigated further to better understand hidden features and biomarkers of prostate lesions with mpMRIs.*

## 1. Introduction

Prostate Cancer is the most common malignancy and the second leading cause of cancer for men in the United States (Silberstein et al., 2013). Thus, it is vital that early and accurate diagnosis of prostate cancer is achieved for appropriate treatment and improved prognosis. Diagnostic tests such as the prostate specific antigen (PSA) serum test and the digital rectal exam (DRE) have been used in the past but have proven to deliver unreliable results. PSA can deliver false positive results since it is also secreted by normal, hyperplastic and malignant tissue. In addition, DRE is an unnecessary invasive procedure that can precipitate pain among other side effects and does not significantly reduce mortality. This ineffectiveness stems from the fact that with DRE, only peripheral zone tumors can be identified in asymptomatic men and sometimes low-grade tumors can even go unnoticed (Ojewola et al., 2013). Ultimately, both tests may lead to a higher chance of false positives and negatives than if another diagnostic test were performed. Such results would negatively affect risk stratification and clinical cancer staging, which are essential to providing the appropriate treatment level.

To better assess the presence of prostate cancer through noninvasive procedures, medical imaging techniques such as multiparametric Magnetic Resonance Imaging (mpMRI) have shown promising results in diagnosis, localization, risk stratification, and staging of clinically significant local prostate cancer (Ghai & Haider, 2015). mpMRIs consist of different scanning methods that develop into four main sets of images: T2-weighted images (T2WI), diffusion-weighted imaging (DWI), dynamic contrast-enhancement (DCE), and MR spectroscopy (Stephenson et al., 2014). However, even though mpMRI has provided a much improved rationale for diagnosis, the analysis and interpretation of these scans through the standardized Prostate Imaging Reporting And Data System (PI-RADS) by radiologists are largely subjective and inconsistent due to varied experience and training in the field of prostate cancer diagnostics. This may result in the increased possibility of false positives/negatives. These inconsistencies may be addressed with computer aided detection (CAD) algorithms to provide a more standardized, objective diagnosis that minimizes the risk of human error (Turkbey & Choyke, 2018).

In the last few years, several machine learning algorithms and pipelines have been proposed for this exact purpose (Litjens et al., 2014; Ehrenberg et al., 2016) and have achieved results that outperform those of trained radiologists. However, the design of such algorithms for feature extraction in mpMRIs require a certain level of domain expertise in order to increase the accuracy of the classification algorithms. With this range of experience in prostate cancer diagnostics, different researchers have reported various algorithmic implementations. Even more recently, the advent of deep convolutional neural network (CNN) techniques has led to great success in natural image processing (Liu et al., 2017; Krizhevsky et al., 2017; Simonyan & Zisserman). For medical imaging, however, the effectiveness of deep learning in a clinical setting is hampered by three major challenges: the heterogeneous raw data, the relatively small sample size classification (Esteva et al., 2017), and the "black box" of CNNs, which is due to their multilayer non-linear structure that is non-transparent and uninterpretable to humans (Hayashi, 2020). In this paper, we address these challenges by incorporating heterogeneous data with proper preprocessing to detect lesions in three zones of the prostate: the peripheral zone (PZ), the transition zone (TZ), and the anterior fibromuscular stroma (AS). Detection pipelines are established for each zone independently as there has been proven to be significant differences in features and signs seen in each zone (Puech et al., 2012). From these features respective to each zone, non-deep learning and deep learning algorithms are compared regarding their performance in accurate lesion detection. Furthermore, important (salient) features in lesion determination from each non-deep learning classifier are determined to better understand the rational for classification. As a supplementary objective, we also hope to better unveil the black box of CNNs by improving the interpretability of results through saliency mapping to highlight the salient features of an image that the CNNs uses to justify its output in lesion detection.

## 2. Literature Review

In the last decade, there have been significant developments in machine learning applications in the medical field, especially in cancer diagnostics. Chan et al. (Chan et al., 2003) were the first to implement a multi-parametric CAD system for the diagnosis of prostate cancer. In their approach, they used line-scan diffusion, T2, and T2-weighted images in combination with an SVM classifier, to identify predefined areas of the peripheral zone of the prostate for the presence of prostate cancer. This study set the foundation for future algorithmic developments in prostate cancer detection and classification using today's traditional machine learning classifiers.

Building on this, Litgens et al. (Litjens et al., 2014) made the first prostate MRI CAD system that was evaluated on a per-patient basis and was compared with the prospective performance of radiologists. This fully automated CAD system included a novel combination of segmentation, voxel classification, candidate extraction, and candidate classification for a more accurate diagnostic outcome compared with the ground truth of a biopsy. Even though performance evaluation showed that it outperformed the state-of-the-art at the time, this comparison had its limitations due to different evaluation data sets involved. To address this issue in our paper, we used the same mpMRI dataset for both non-deep learning classifiers and deep learning classifiers so that inputs were standardized and allowed for valid comparison between classifiers.

Focusing more on deep learning developments, in a 2019 study, collaborators developed a deep-learning system (DLS) for the prostate cancer grading scale (Gleason scoring) based on whole tissue images (201, 2019). This DLS achieved a diagnostic accuracy of 0.7 (scale of 0.5 (random) to 1). This was an impressive result when compared with 29 expert pathologists' diagnostic accuracy of only 0.61. Furthermore, follow-up data confirmed a better patient risk stratification by the DLS. Overall, this study demonstrates how machine learning can improve well established standards such as the Gleason scoring by eliminating subjective evaluation by the human eye, thus yielding to more precise prognostication. While this study used whole tissue samples as an input, we used mpMRIs instead of a biopsy for noninvasive diagnosis to increase patient comfort and to hopefully yield similar results. Finally, Liu et al. (Liu et al., 2017) constructed a accurate CNN model, Xmasnet, to detect prostate lesions using an image of the entire prostate. To instead determine the performance of CNNs in each zone, we created our own CNN architecture (Figure 1) inspired by Xmasnet to fit our smaller input images.

## 3. Purpose

1. Extract statistical and textural features from mpMRIs in separate zones of the prostate.

2. Implementation and analysis of machine learning classifiers to determine the best classifier in each zone.

3. Increase interpretability of CNN results through saliency mapping.

## 4. Methods

### 4.1. Data Set

All data used for this paper was provided by the 2017 PROSTATEx Challenge held in conjunction with the 2017

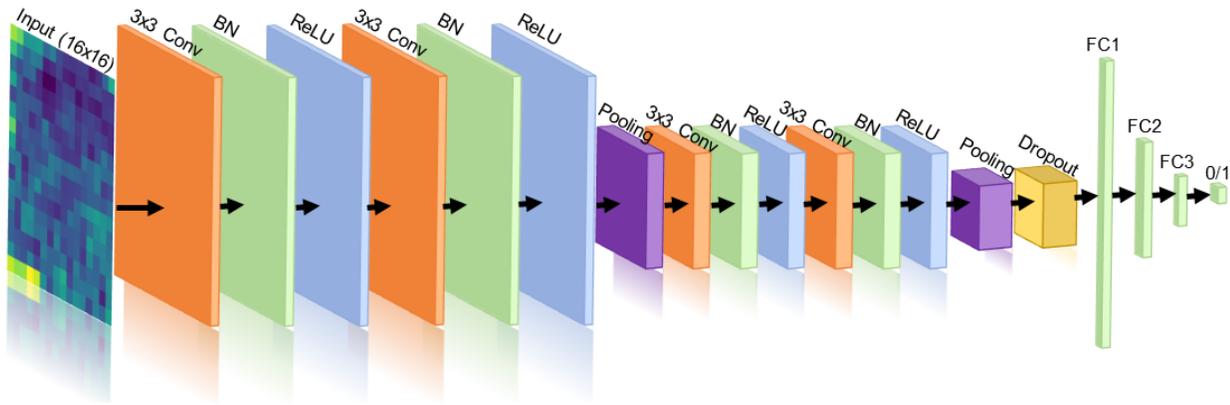

*Figure 1.* Custom CNN Architecture. Conv: convolutional layer; BN: batch normalization layer; ReLU: rectified linear unit; Pooling: max pooling layer; FC: fully connected layer.

SPIE Medical Imaging Symposium. This collection is a retrospective set of prostate MR studies. Studies include T2-weighted (T2W), diffusion-weighted (DW) imaging, and dynamic contrast-enhanced (DCE).

**T2-weighted imaging** Given a proton from a hydrogen atom, whenever the magnetic dipole moment is placed into an external magnetic field, it will feel a torque that will tend to align it with the external magnetic field. The magnetic dipole moments will try its best to align with the magnetic field lines, but it will not line up exactly. Instead, it will precess around on an axis, which can be either spin-up or spin-down (with spin-down having more energy). In this case, the external magnetic field is the electromagnetic (EM) pulse given by the MRI machine to give spin-up protons enough energy to become spin-down protons. Once the EM pulse is turned off, the excited proton returns to equilibrium and gives off photons which the MRI machine can measure. While the amount and locations of photons are measured in T1 imaging, T2-weighted images are created based on T2 relaxation times, which is the time taken for spinning protons to lose phase coherence among other proton-containing nuclei. Since T2-weighted imaging assigns pixel intensities based on the density of hydrogen atoms in an area, it can map relative water content within the body to create recognizable structures and tissues. Within the prostate, the peripheral zone has the highest water content and is thus the brightest part of the image.

**Diffusion-weighted imaging** This is a form of MR imaging based upon measuring the random Brownian motion of water molecules within a voxel of tissue. In general, highly cellular tissues or those with cellular swelling exhibit lower diffusion coefficients, which are represented as lower intensity values in an apparent diffusion coefficient map. This type of imaging is particularly useful for determining tumor presence as more tissue results in lower diffusion of water due to the restriction of space, leading to lower signal intensities.

**Dynamic contrast-enhanced imaging** This sometimes also referred to as permeability MRI and is one of the main MRI perfusion techniques which calculates perfusion parameters by evaluating T1 shortening induced by a gadolinium-based contrast bolus passing through tissue. The most commonly calculated parameter is k-trans and is the one given in this dataset. Ktrans is measure of capillary permeability and has higher values (corresponding with higher pixel intensities) for areas with more blood flow. This is a significant tool when it comes to detecting tumors, especially those that have induced angiogenesis to get more blood flow to the tumor.

The images were acquired on two different types of Siemens 3T MR scanners, the MAGNETOM Trio and Skyra. T2-weighted images were acquired using a turbo spin echo sequence and had a resolution of around 0.5 mm in-plane and a slice thickness of 3.6 mm. The DCE time series were acquired using a 3-D turbo flash gradient echo sequence with a resolution of around 1.5 mm in-plane, a slice thickness of 4 mm and a temporal resolution of 3.5 s. Finally, the DWI series were acquired with a single-shot echo planar imaging sequence with a resolution of 2 mm in-plane and 3.6 mm slice thickness and with diffusion-encoding gradients in three directions. Three b-values were acquired (50, 400, and 800), and subsequently, the apparent diffusion coefficient (ADC) map was calculated by the scanner software. All images were acquired without an endorectal coil (Litjens et al., 2014; 2017; Clark et al., 2013). For our paper, we mainly focused on using T2-weighted images and ADC maps for feature extraction, feature extraction, and the resulting binary classification for lesion determination.

## 4.2. Data Storage

The raw data provided included the acquired MR images encoded in DICOM format while Ktrans images were provided in MHD format. Ktrans is a key pharmacokinetic parameter computed from the available Dynamic contrast-enhanced T1-weighted series. Each patient had one corresponding Ktrans image but can have up to three lesions, so detection was evaluated on a per-lesion basis. The annotations for the mpMRIs were provided in csv files that contained information such as the de-identified patient number (compliant with HIPAA rules), a finding number corresponding to each lesion found in a single patient, a centroid location marked by a radiologist (biopsied for the ground truth), and a DICOM description to match the annotation to the image data. The overall data was already separated into train and test sets by the image repository with a total of 327 lesions in the train set and 206 in the test set. All data was sorted through and compiled automatically through scripts written in-house. The lesion pixel arrays and corresponding annotations were stored in an hierarchical data format version 5 (HDF5) for efficient storage and querying of data. This storage method allowed for almost instantaneous speeds of data extraction from the raw data and significantly reduced runtime when querying the data (Folk et al., 2011).

## 4.3. Preprocessing

We recognized that images which contained similar objects or scenes often had different intensity ranges that made it difficult to compare them manually, especially in T2WIs. Since the focus of this study was to primarily use T2-weighted images (T2WI) and ADC map images, we decided to apply an intensity range standardization algorithm provided by Medpy.filter to transform T2WI image intensity ranges to a common standard intensity space without any loss of information. This was done with a multi-segment linear transformation model. Thus, this method allowed for valid comparison of prostates without signal intensity bias created from different scanning conditions. This algorithm works by defining a standard intensity space through an intensity value range. During the training phase with the T2WIs in the training set, the intensity values at certain cut-off percentiles of each image were computed and a single-segment linear mapping from them to the standard intensity space range limits were created. Then the images' intensity values at several landmark percentiles were extracted and passed to the linear mapping to be transferred roughly to the standard intensity space. The mean of all these mapped landmark intensities formed the model learned. When the model was applied on a new test image, the image's intensity values were extracted at the cut-off percentile as well as at the landmark percentile positions. This resulted in several segments. Using these segments, the corresponding standard intensity space range values, and the learned mean landmark values, a multi-segment linear transformation model was created for the image. This was then applied to the image's intensity values to map them to the standard intensity space, creating the new, intensity-standardized image (Figure 2), ready for feature extraction (Nyul et al., 2000). This signal intensity range standardization algorithm was not applied to ADC map images as ADC is a quantitative measurement (Bonekamp et al., 2018).

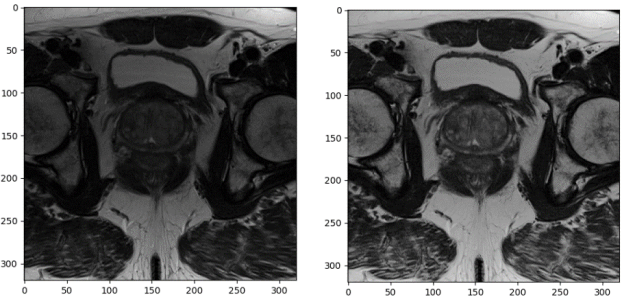

*Figure 2.* Before and after images of Intensity Range Standardization.

## 4.4. Data Augmentation

To split the data, the number of lesions from each zone were counted. In the training set, there was found to be 188 peripheral zone (PZ) lesions, 82 transition zone (TZ) lesions, 55 anterior fibromuscular stroma (AS) lesions, and 2 seminal vesicle (SV) lesions. In the test set, there were 113 PZ lesions, 59 TZ lesions, 34 AS lesions, and 2 SV lesions. Since there were no malignant SV lesions for our model to learn from, we excluded SV lesions from our study, similar to the procedure in Chen et al. (Chen et al., 2019). This was a part of our study that we hoped other studies could address given a more inclusive data set.

The main limitation for the provided data set was that the test set's ground truth was not given (still used in an ongoing challenge). The original purpose of that challenge was to classify lesions to an appropriate clinical significance score, in which clinical significance was defined as having a Gleason score of greater than 7. Knowing that all lesions contained some lesion grade, all images under a zone label were denoted to have a lesion (truth label of '1'), denoting that they were lesions. Negative labels (truth label of '0') for lesions were manually extracted from patient prostates with no noted lesions noted for a zone, meaning that there was no lesion there, thus preventing data leakage. Within both training and test sets, the T2WI and ADC image sets were split into 3 zones: PZ, TZ, and AS. To generate the samples needed for feature extraction, a 16x16 pixel area for T2WI and 6x6 pixel area for ADC images were cut out from the centroid of the positive lesion marked by the radiologist. To generate the same size areas from negative

MRI samples, random sampling was done at the manually bounded area until the positive and negative ratio was 1 to 3 (25% positive, 75% negative) for each zone. In addition, during the sampling process, T2WI and ADC images were paired per location so that the features extracted would be relevant for a single location towards binary classification of the possible lesion in the experimental stage. A summary for sample quantities is recorded in Table 1.

### 4.5. Feature Extraction

For each zone and MRI type combination, 13 features were extracted, totalling 26 features for each sample. First order statistics based on the pixel signal intensities were extracted such as the 10th percentile and average intensity (Peng et al., 2013). Second order texture features were extracted based on Tamura and Haralick features. These are considered second order with respect to how they consider the relationship between groups of two pixels in an image.

The most common texture analysis for image classification is based on the gray-level co-occurrence matrix (GLCM) by Haralick et al. (Haralick et al., 1973). The GLCM is a matrix that is defined over an image to be the distribution of co-occurring pixel values (gray level intensity) at a given offset. This matrix is then used to make various texture measure calculations. In our case, we automatically calculated the GLCM matrix for each lesion MRI sample with an offset of one pixel and zero degrees as the direction for the neighbor pixel. We originally experimented with multiple degrees and then summing the output values together, but the classification results did not show significant improvement from only calculating zero degrees.

There are three main categories of Haralick features totaling to 13 features, of which we used six. In the first category, the contrast group uses weights related to the distance from the GLCM diagonal (weights on the diagonal show no contrast) to calculate a weighted average value as an output. The contrast group includes the contrast value (sum of squares variance) where weights increase exponentially, dissimilarity where weights increase linearly, and homogeneity (inverse difference moment) where weights decrease exponentially from the diagonal. In the second category, the orderliness group features define how regular pixel values differences are within a preset frame. The weights to calculate the weighted average value are based on how many times a given pair occurs. The orderliness group includes the angular second moment (ASM) and energy which have high values when the frame contents are very orderly. In the final category, descriptive statistics of the GLCM matrix are synthesized such as correlation, which measures the linear dependency of gray levels on those of neighboring pixels. In summary, we used GLCM texture features (formulas in Table 2) of contrast, dissimilarity, homogeneity, ASM, energy, and correlation to be incorporated into each sample's feature set.

Tamura et al. (Tamura et al., 1978) presented that six textural characteristics had high correlation to human visual perception, which were coarseness, contrast, directionality, line-likeness, regularity, and roughness. Of these six, we will use coarseness, contrast, and roughness as features (formulas in Table 2). Coarseness in this case refers to the distances of notable spatial variations of grey levels which implicity describes the size of primitive elements (texels) forming a texture. Contrast measures how much gray levels vary in an image and to what extent their distribution is biased towards black or white. The final roughness feature is given by summing coarseness and contrast values.

Finally, we extracted third order statistical features of skewness and kurtosis. Skewness measures the lack of symmetry, which is medically significant as a heavy bias toward lower values is associated with the presence of a lesion, especially in ADC maps. Kurtosis measures whether the data is heavy-tailed or light-tailed compared to a normal distribution. This can help us determine the impact of outlier signal intensities within a sample. Once a feature was calculated for each lesion, the feature vector was appended as another column to an existing feature matrix and used as input to classifiers.

### 4.6. Non-Deep Learning Classifiers

For our method, each zone had an independent classification pipeline, meaning that each classifier did not incorporate any data between different zones. For this section, we evaluated the performance of non-deep learning classifiers such as Logistic Regression, Support Vector Machine (SVM), Random Forest, and XGBoost. The results of these classifiers were compared with each other and with the deep learning classifier. Except for XGBoost (Chen & Guestrin, 2016), all non-deep learning classifier algorithms were provided by the python library Scikit-learn (Pedregosa et al.).

**Logistic Regression** A logistic regression model is formulated mathematically by relating the probability of some event conditional on the vector of explanatory variables to attain a binary output at a given threshold value. Some of the variables may be useful for the classification, but when combined with other variables, they may be ineffective if they are highly correlated with other variables. Variables that contribute redundant information can be omitted so that only those variables with the best diagnostic accuracy are retained for the model. For our purposes, a L1/Lasso regularization method was first applied as a feature selection technique to select for significant features with non-zero coefficients to decrease noise and to be used in classification. Then, a 10-fold cross validation was performed on training zone data for hyperparameter tuning. In the end, the logistic regression algorithm used was set to have a L1 penalty and liblinear solver. This setting was used for all zones as a baseline classifier.

|  | Zone Type | | |
| --- | --- | --- | --- |
| Data Set | PZ | TZ | AS |
| Train | 188/748 | 82/322 | 55/215 |
| Test | 113/453 | 59/239 | 34/132 |

Table 1. Data augmented so that there were 25% positive, 75% negative samples in each zone. Number of lesions were the same in T2WI & ADC images.

| Texture Feature | Formula |
| --- | --- |
| Haralick Angular Second Moment | $\sum_i \sum_j p(i,j)^2$ |
| Haralick Contrast | $\sum_i \sum_j (i-j)^2 p_d(i,j)$ |
| Haralick Correlation | $\frac{\sum_i \sum_j (ij)p(i,j) - \mu_x \mu_y}{\sigma_x \sigma_y}$ |
| Haralick Dissimilarity | $\sum_{i,j=0} P_{i,j} |i-j|$ |
| Haralick Energy | $\sqrt{ASM}$ |
| Haralick Homogeneity | $\sum_i \sum_j \frac{1}{1+(i-\mu)^2} p(i,j)$ |
| Tamura Coarseness | $\sum_{i=x-2^{k-1}}^{x+2^{k-1}-1} \sum_{j=y-2^{k-1}}^{y+2^{k-1}-1} f(i,j)/2^{2k}$ |
| Tamura Contrast | $\sigma/(\alpha_4)^n$ where $\alpha_4 = \mu_4/\sigma^4$ |
| Tamura Roughness | Coarseness + Contrast |

Table 2. Formulas to calculate Tamura and Haralick features from a Gray-Level Co-occurrence Matrix (GLCM).

**Support Vector Machine** An SVM generates maximal margin hyperplanes which separates the data into different groups. If the data is not linearly separable, then a kernel function is used to move the data into a higher dimension so that separation may be easier in that space. Since the number of observations were greater than the number of features for our data, a nonlinear Gaussian Radial Basis Function (RBF) kernel was used with a C value of 0.05. This setting was used for all zones.

**Random Forest** The Random Forest algorithm consists of many individual decision trees that operate as an ensemble. Each individual tree in the random forest produces a class prediction and the class with the most votes becomes our model's prediction. The main advantage with using random forests is that many relatively uncorrelated models (trees) operating as a committee outperforms any of the individual constituent models. While each deep tree in the random forest tends to overfit (leading to low bias and high variance), averaging relatively uncorrelated high variance models via random forest produces a much lower variance ensemble to optimize the bias-variance tradeoff. The best hyperparameters were determined manually so that the function criterion to measure the quality of a split was 'entropy' and the number of trees in the forest were 1000. Entropy here refers to criterion for calculating information gain, which decision trees in the random forest use to split a node and determine the structure of the tree in a way that reduces entropy. This setting was used for all zones.

**XGBoost** XGBoost is an implementation of a gradient boosted tree algorithm. Gradient boosting refers to the sequence of classification models that are learned from the data in which each classifier gives higher weight to incorrectly classified instances. This weight allows the next classifier in the sequence to more likely sample those instances with higher weights. In contrast to random forest, sequential methods such as XGBoost are used to train an ensemble of high bias models (shallow trees in this case) to increase the variance and decrease the bias. Due to the nature of this ensemble, this technique is highly prone to overfitting, but the learning rate (eta), gamma, and subsampling parameters can be adjusted to control overfitting. During hyperparameter tuning, A few other parameters were also adjusted. The Colsample_bytree parameter is the subsample ratio of columns when constructing each tree. Max depth is the maximum depth of the tree with increased depth leading to increased complexity and chance of overfitting. Subsample is the subsample ratio for the training instances used, in which lower values signifies that XGBoost would randomly sample less of the training data prior to growing trees, preventing overfitting. Finally, n_estimators is the number of trees used. To automatically tune hyperparameters on the training set, three-fold cross validation was performed and Scikit-learn's RandomizedSearchCV was used to return the best hyperparameters for each independent zone. Refer to Table 3 for the hyperparameter values set for each zone.

### 4.7. Deep Learning Classifier

A deep learning convolutional neural net (CNN) was also established for each zone. The CNN assigns importance (learnable weights and biases) to various image features in the image to be able to detect the presence of a lesion. In this paper, our architecture (Figure 1) was composed of various layers with an input of 16x16 T2WI MRIs as ADC images were too small to be used as input. Convolutional layers were the core part of this model as they allowed the the neural net to essentially extract features from an input image through many filters. Then, the output of the convolutional layer was processed by a batch normalization (BN) layer. This was important as it ensured that each layer saw the same distribution of inputs from the previous layer at each iteration of training, thus learning happened at faster rate as the model did not also have to deal with distributional (covariate) shifts between iterations. After this layer, a nonlinear Rectified Linear Unit (ReLU) activation function was used to process the basic outputs of the neuron and allowed for complex mappings of outputs to the next layer of the architecture. The purpose of specifically using ReLU was to provide both a nonlinearity in addition to a constant gradient (unlike a tanh activation function). After this, a max pooling layer was used to reduce the number of extracted features and to avoid overfitting. In addition to BN layers, dropout layers also prevented overfitting and improved training time, albeit through a different mechanism. Thus, we decided to experiment with them and concluded with slightly improved results in doing so (Garbin et al., 2020). Finally, we included an Adam optimizer and a cross entropy loss function as defined by Xmasnet (Liu et al., 2017).

## 5. Results

The performance of each classifier in its respective zone was evaluated through a Receiver Operating Charactistic (ROC) curve, Precision Recall Curve (PRC), and F1 score. The ROC curve is defined as a plot of true-positive ratio (TPR) against false-positive ratio (FPR) when the threshold c moves on a real number line. The area under the ROC curve (AUC) was taken as a measure of separability for binary lesion classification. However, since the data set was imbalanced with 25% positives and 75% negatives, this was accounted for by using PRCs and F1 scores, which do not introduce true negative results in their calculation. This allowed for the accuracy of measuring true positives to be better interpreted compared with the AUC of a ROC curve. The F1 score is determined by calculating the harmonic mean between precision (positive predictive value) and recall (sensitivity) to combine these two statistics into one value. The PRC plots the precision and recall at different threshold values into a graphical form so that the AUC can be taken. ROC and PRC curves with their corresponding AUC scores and F1 scores can be found in Appendices A.1 to A.3.

### 5.1. Salient Features for Non-Deep Learning Classifiers

After the best classification result was achieved on the test set of a zone, the coefficient vector for each classifier (except for SVM) was used to generate a bar graph and salient features used for classification were reported. A coefficient vector could not be taken from SVM because a nonlinear kernel was used, and the mapping function could not be analytically determined from an infinite-dimensional transformed space.

**Logistic regression** The logistic regression algorithm mainly used second and third order statistical features. Specifically, there was a high preference for texture features such as Haralick correlation and Tamura coarseness and third order statistics such as skewness and kurtosis. There was not a preference for either T2WI or ADC maps as both were highly prioritized for their respective features.

**Random forest** The random forest algorithm performed the best using first and second order statistical features. Specifically, there was a high preference for 10% and average intensity values and second order texture features such as Haralick dissimilarity and contrast. There was not a preference for either MRI type as both were highly prioritized for the features mentioned.

**XGBoost** The XGBoost algorithm utilized all orders of features to derive the best performing model. From ADC images, 10% intensity values were found to salient as were Haralick energy and ASM. From T2WIs, Haralick correlation and Tamura roughness were found to be salient. From both MRI types, features included average signal intensity and skewness.

### 5.2. Salient Features for each Zone

There were various trends noted for which salient features corresponded to each zone to attain an accurate classification.

|  | Hyperparameters | | | | | |
| --- | --- | --- | --- | --- | --- | --- |
| Zone | Colsample_bytree | Gamma | Eta | Max depth | N estimators | Subsamples |
| PZ | 0.73 | 0.009 | 0.058 | 4 | 122 | 0.63 |
| TZ | 0.70 | 0.255 | 0.155 | 2 | 132 | 0.65 |
| AS | 0.71 | 0.013 | 0.143 | 2 | 117 | 0.99 |

*Table 3.* XGBoost hyperparameters were tuned using three-fold cross validation in each zone: peripheral zone (PZ), transition zone (TZ), and anterior fibromuscular stroma (AS).

**Peripheral zone**   Features that were salient for PZ lesions included first order statistical features such as 10% and average intensity as well as second order texture features such as Tamura courseness and Haralick energy and dissimilarity. There was not a significant preference for either MRI type, except for Haralick features where ADC maps were more significant.

**Transition zone**   Features that were salient for TZ lesions included ADC 10% and second order texture features such as Tamura roughness and Haralick, correlation, contrast, and dissimilarity. There was no preference for either MRI type for texture features.

**Anterior fibromuscular stroma zone**   Features that were salient for AS lesions included first order features such as 10% and average intensity and third order features such as skewness and kurtosis. There was no preference for either MRI for salient features.

Throughout all the features, 10% and average intensity as well as skewness contributed the most information for classification, regardless of MRI type.

## 6. Discussion

We will discuss the experimental results from comparing non-deep learning and deep learning classifiers for prostate lesion detection with respect to the region of the prostate that samples were extracted from.

It has been reported that a radiologist following the Prostate Imaging-Reporting and Data System (PI-RADS) to detect lesions from mpMRIs can achieve an AUC of 0.81 to 0.84 (Kasel-Seibert et al., 2016; Junker et al., 2013; Polanec et al., 2016). It seems that a few base classifiers without any hyperparameter tuning can attain a similar perfomance. While the radiologist-like performance is already satisfactory, we believe that there is more room to improve. After extensive hyperparameter tuning on training sets, the models from cross validation transferred over relatively well to the test set and achieved AUC values that were significantly better than the baseline performance. It seems that some non-deep learning classifiers such as logistic regression, random forest, and XGBoost were able to achieve a higher accuracy through hyperparameter tuning. Overall, Xgboost performed similar if not better compared with Random Forest. Logistic Regression with L1 regularization was slightly worse, but still performed better than a well-trained radiologist would in lesion detection. Finally, SVM with a radial basis function (RBF) kernel performed with metrics close to random chance. The poor perfomance of SVM can be attributed to having many heterogenous and independent features as input. The RBF kernel gives equal weighting to noisy and informative features, thus decreasing the performance of the model. However, SVM has proven to achieve great results in prostate lesion detection (Artan et al., 2010; Wang et al., 2017). Hopefully, future studies may augment the data used in this study in such a way to decrease noise and increase the performance of SVM. Even with optimal tuning, these classifiers achieved various results in different zones, which can be explained by the increased medical significance of some features over others.

**Peripheral zone**   For lesions present in the peripheral zone (PZ), first and second order features have the most medical significance for lesion determination as round or ill-defined low-signal intensity masses are an indication of a lesion in both ADC and T2WI MRIs (Hricak et al., 1983). Healthy peripheral zone tissue gives off the highest signal intensity out of all regions owing to its high-water content. Thus, it is relatively easier to diagnose lesions in this zone. Generally, there is also a higher rate of interobserver agreement for PZ lesions as reported by Schoots et al. (Schoots, 2018), which is reflected by the highest performance measurements of the area under ROC and PRC for classifiers in this zone. In regards to the best classifiers of this zone, ensemble algorithms such as random forest and xgboost performed significantly better than others as it combines multiple models together and delivers superior prediction power. Surprisingly, F1 scores were significantly lower than area under PRC throughout all the zones, especially in PZ. We discovered that models tend to classify more lesions as positives as evidenced by the high recall and low precision for positive results. While misclassification is present, from a medical standpoint, it is much more beneficial to detect a possible lesion as a false positive result than to completely miss the lesion as a false negative.

**Transition zone** For lesions present in the transition zone (TZ), first order statistics and second order texture features influenced classification the most. The pixel intensities found in this zone have a more heterogenous distribution for healthy tissue, emphasizing the importance of texture analysis. Lesions in TZ appear often as a homogenous mass possessing ill-defined edges with lenticular or "water-drop" shapes on a T2WI (Akin et al., 2006; Barentsz et al., 2012). However, not all lesions are obvious in a heterogenous background, making the classification problem more difficult than in PZ. The classifiers' performances reflect this issue, having the lowest classifier performance metrics of all zones. Nonetheless, XGBoost performed significantly better than random forest in this zone compared with other zones. This is likely due to the correlated texture features having a beneficial, correctional effect during the iterative training process in XGBoost. For TZ lesion detection, we believe that more texture features could be incorporated in future studies to give more information to a model to train on. However, proper regularization techniques would still need to be utilized to reduce noise.

**Anterior fibromuscular stroma zone** For lesions present in the anterior fibromuscular stroma zone (AS), first and third order statistical features influenced classification the most. Lesions in this area present with a significant hypointensity, lower than that typically expected for other zones (Yu et al., 2017). Thus, it is understandable that texture features are not as salient and features that directly relate to signal intensity are much more important in classification. Most importantly, the deep learning CNN performed the best in AS and overall as it was not limited by the pre-extracted features provided to all non-deep learning models. It was able to learn its own feature set and classify based off of that. To interpret the salient features the CNN used in classification, a saliency map was generated to explain the high performing classifier (Figure 3). From this image, we can understand that the model clearly recognizes the shape and hypointensity of lesions found in the AS zone. Other than deep learning CNNs, performance in this zone is mid-level with metric values mostly falling between those of other zones. Another surprising finding is that random forest performed better than XGBoost with respect to resulting metrics. This is likely due to salient feature independence that benefits the independence of decision trees within the random forest in contrast to what was seen in TZ. Another important result to note is that the area under the ROC curve for logistic regression is significantly worse than that of other zones, demonstrating the lowered ability to differentiate between positives and negatives. However, it also has a significantly higher area under the PRC than other zones, indicating the increased accuracy of predicting true positives. This means that if the logistic regression classifier labels a lesion as positive, it has a higher chance of being a true positive compared to the same classifier found in other zones.

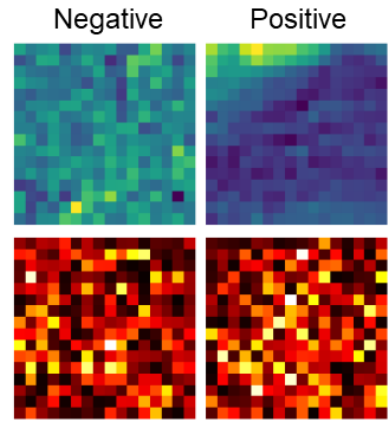

*Figure 3.* Saliency Map from AS zone to interpret the high performance of the deep learning model. Bottom images are the saliency maps derived from the top T2WI images, in which orange to white colors indicate high gradients while darker colors indicate low gradients. Based on the gradients, we can determine which pixels contributed most to the CNN's decision of classification.

## 7. Conclusion and Future Work

In this study, we uncovered the best type of classifier and salient features for each zone of the prostate. For the PZ and TZ zones, ensemble algorithms performed the best while the deep learning CNN soared in performance in the AS zone. We also discovered that the PZ and AS zones were more dependent on statistical features while the TZ zone was more dependent on textural features. The development of CAD systems for prostate lesion detection holds to be the most promising noninvasive imaging diagnostic rationale which leads to high accuracy of detection and minimizes human error. This would allow for medical professionals to promptly begin treatment after lesion detection and grading to deliver the appropriate level of treatment for a patient to end in a favorable prognosis. While the results were encouraging, we believe that classification models used in this study can be improved with more diverse samples from other data sets to not only validate the results of this study but also improve the performance of these models with more rigorous training. Another limitation that prevented better

CNN classification was that this model was not pre-trained on existing images, but rather an original creation inspired by other models and trained on the augmented images within this dataset. If more datasets can be acquired and augmented in a similar way to provide more samples for the CNN to train on, then the performance of the CNN will be improved for more reliable and accurate prostate lesion detection as a CAD system.

Since we determined salient features on a per zone basis, future studies may utilize these indicators to build reliable and accurate supervised machine learning classifiers for whole prostate lesion detection to avoid the blackbox of CNNs altogether. We hope that our work will provide a more standardized basis regarding which algorithms work best for each zone of the prostate and provide future direction in investigation of improving the types of machine learning classifiers highlighted for implementation of CAD systems.

## 8. Acknowledgements

I would like to offer my special thanks to the SSI staff for providing this amazing opportunity to learn advanced concepts and apply them in my first time in research. Their willingness to sacrifice their summer to organize this science institute is very much appreciated. I am particularly grateful for the assistance given by my mentor, Nithin Buduma, for helping me brainstorm ideas, overcome any technical issues I had with my dataset or code, and explain high level concepts for my understanding and implementation in this paper. Finally, I would like to thank my lab group for supporting my ideas, delivering helpful feedback on my presentations, and asking insightful questions.

# A. Appendix

## A.1. Peripheral Zone Graphs

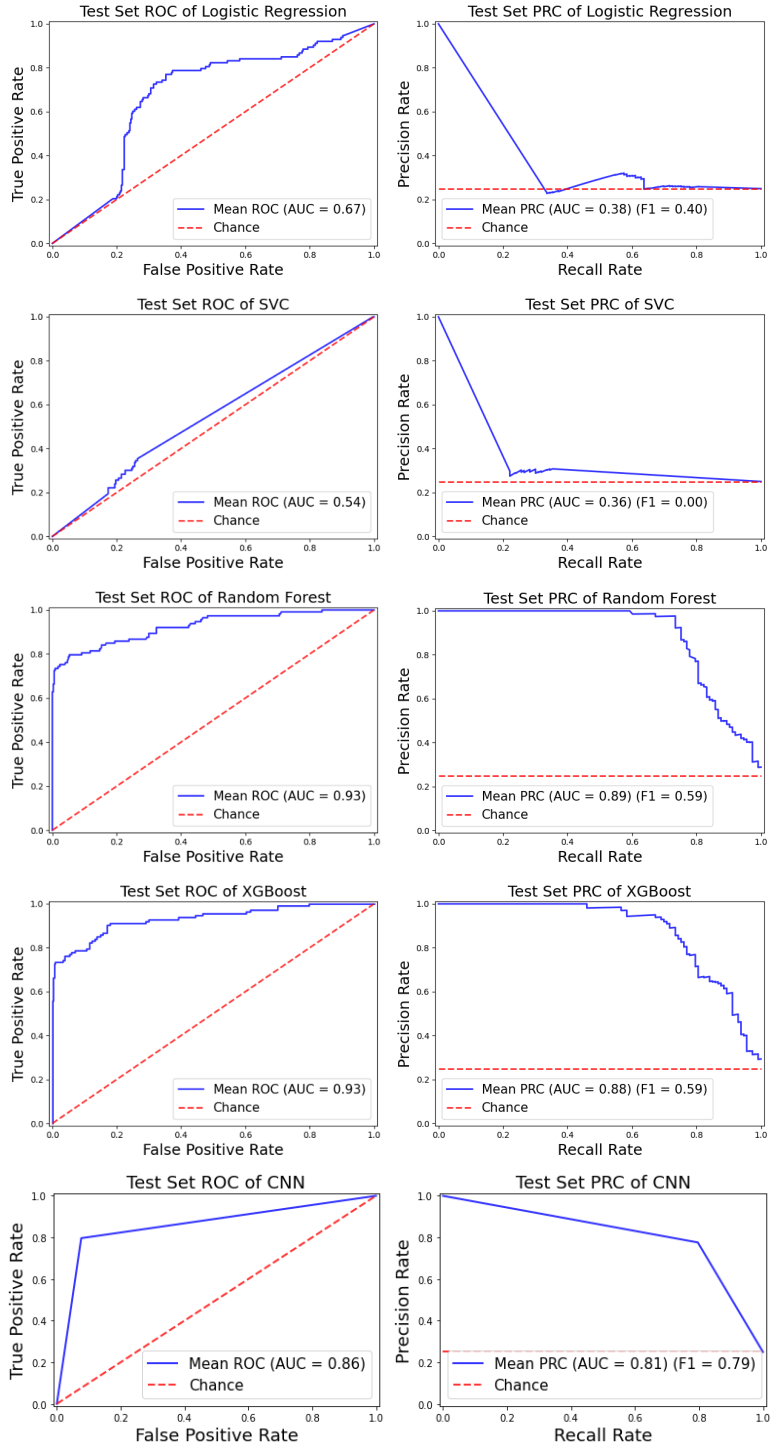

## A.2. Transition Zone Graphs

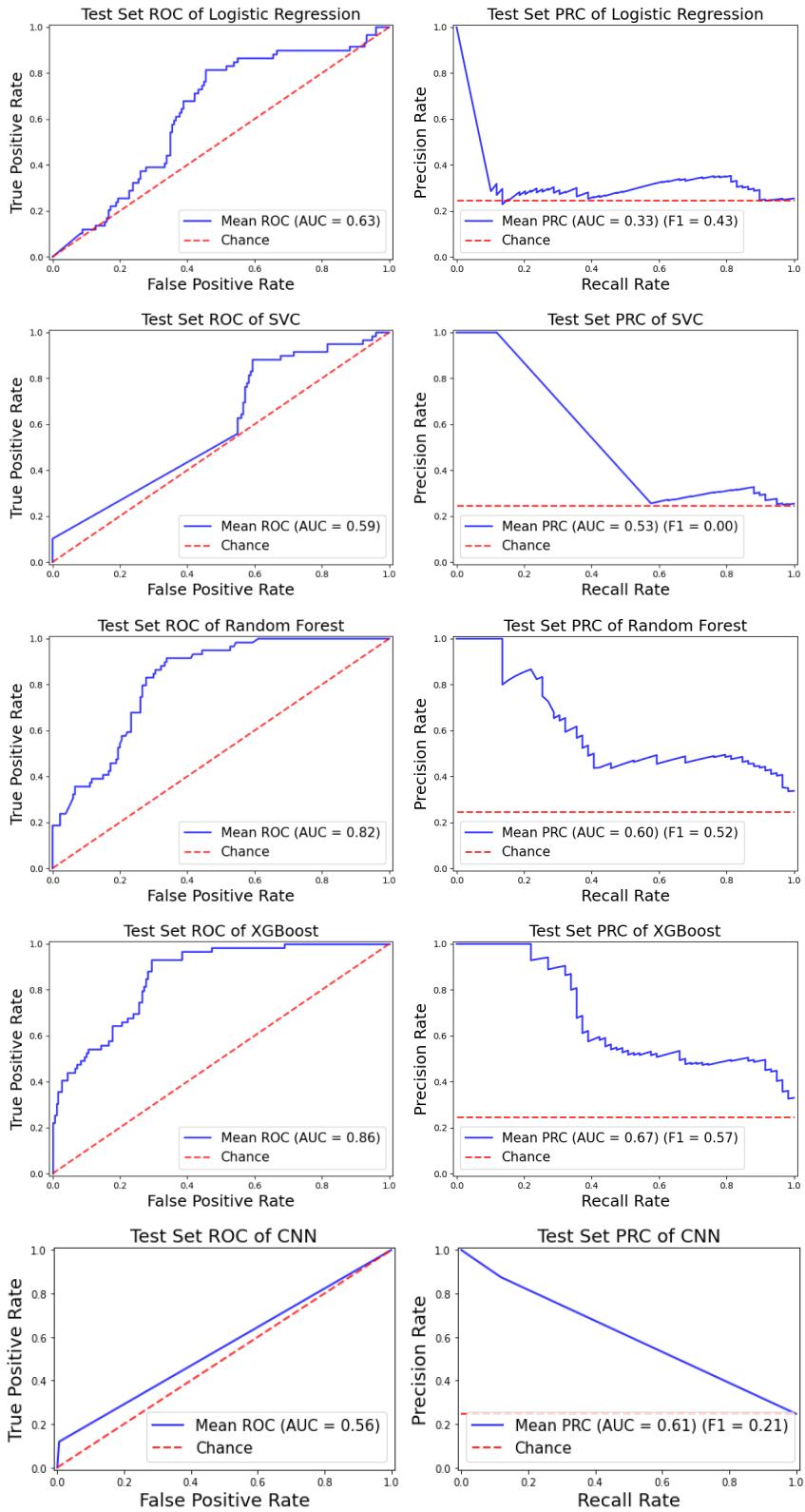

## A.3. Anterior Fibromuscular Stroma Zone Graphs

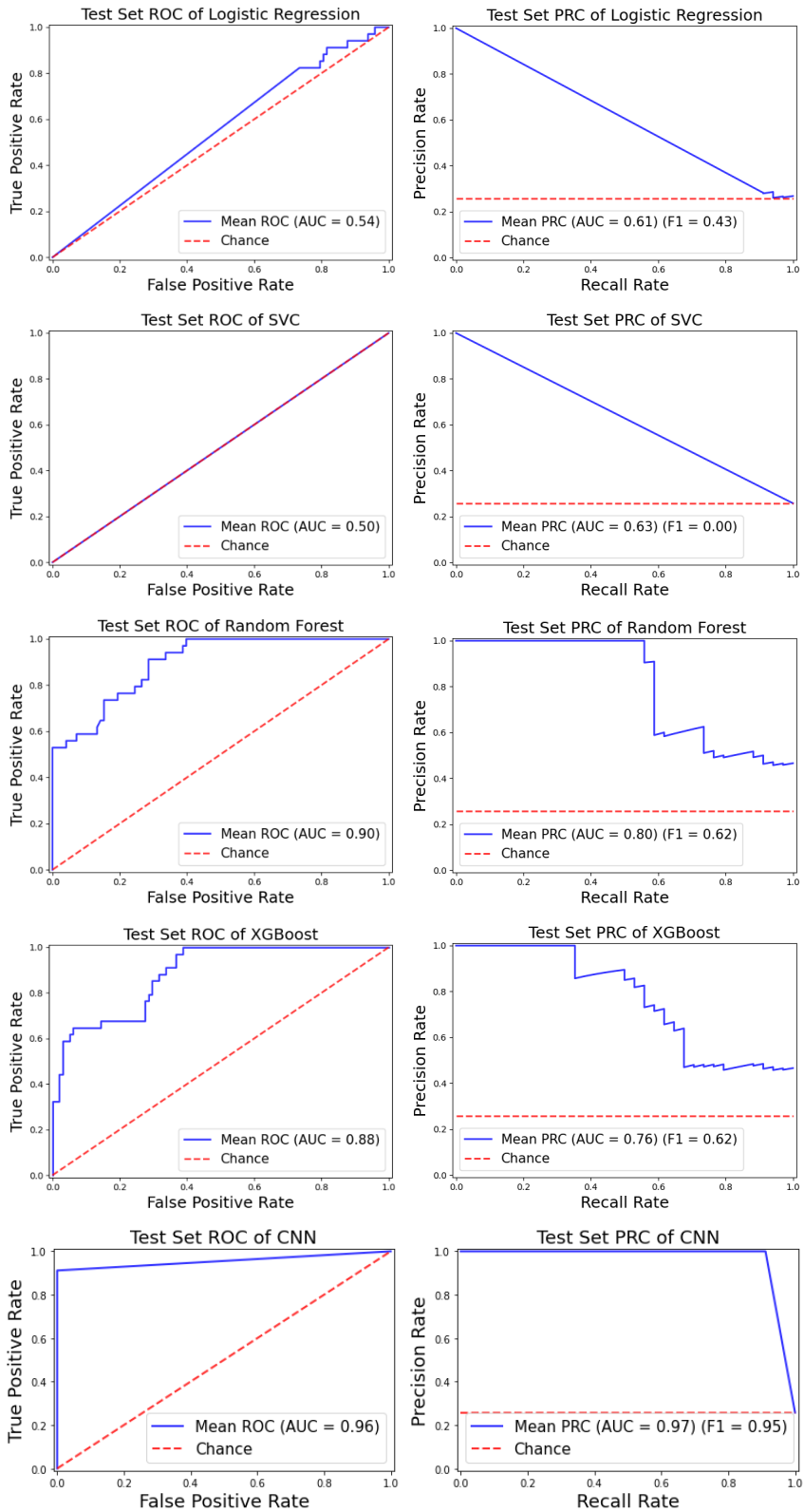

# Prostate Lesion Detection and Salient Feature Assessment Using Zone-Based Classifiers

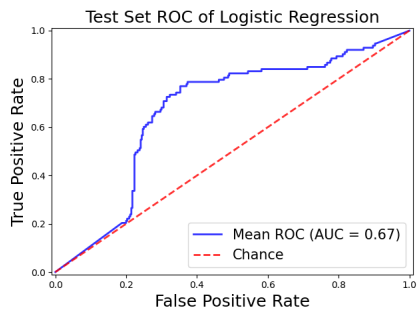
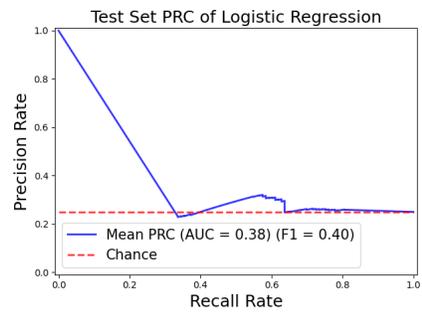
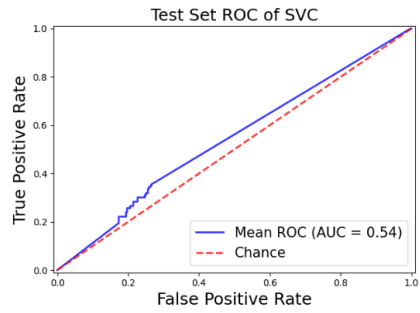